\documentclass{aastex701}

\usepackage{amsmath}
\usepackage{multirow}
\usepackage{booktabs}
\usepackage{tabularx}
\usepackage{threeparttable}
\usepackage{siunitx}
\usepackage[version=4]{mhchem}
\usepackage{subcaption}

\shorttitle{GeV $\gamma$-ray emission in S140}
\shortauthors{Yang et al.}

\begin{document}

\title{GeV $\gamma$-ray emission in the star-forming region S140}

\author{Li-Nuo Yang}
\affiliation{School of Physics and Astronomy, Sun Yat-sen University, Zhuhai 519082, China}
\email{yangln26@mail2.sysu.edu.cn}   

\author{Pak-Hin Thomas Tam}
\affiliation{School of Physics and Astronomy, Sun Yat-sen University, Zhuhai 519082, China}
\affiliation{CSST Science Center for the Guangdong-Hong Kong-Macau Greater Bay Area, Sun Yat-Sen University, Zhuhai 519082, China}
\email[show]{tanbxuan@sysu.edu.cn}
\correspondingauthor{Pak-Hin Thomas Tam}

\begin{abstract}
We report the detection of a GeV $\gamma$-ray source in the 
S140 HII region using 17.75 years of \textit{Fermi}-LAT data, with a 
source-detection significance of $\sim32.1\sigma$ in the 0.1--500~GeV energy band. 
The emission is significantly extended (AIC evidence ratio 
above 1~GeV is $2\times10^{9}$) and is therefore designated as 
4FGL~J2220.8+6319e. 
The $\gamma$-ray emission coincides spatially with a 
molecular cloud that hosts the massive young stellar 
object S140~IRS1. 
Its spectrum is best fitted by a log-parabolic spectrum 
($TS_{\mathrm{cur}}(\mathrm{LP}) = 104.1$) and can be reproduced by both hadronic 
and leptonic models. 
The jet driven by S140~IRS1 offers a viable particle acceleration site, yielding 
injection timescales and maximum particle energies consistent with theoretical 
expectations, whereas stellar winds are energetically disfavored.

\end{abstract}


\section{Introduction}
\label{intro}
Young stellar objects (YSOs) typically form via the gravitational collapse of a dense parent molecular cloud. Among them, protostars with masses exceeding $8 M_{\odot}$ are classified as massive young stellar objects (MYSOs). The formation of MYSOs is often accompanied by intense accretion and subsequent vigorous jets, with terminal velocities reaching $300$ to $1500$ km s$^{-1}$~\citep{2021MNRAS.504.2405A}. While MYSO jets are known to emit from radio to optical wavelengths, their potential ability to emit high-energy radiations remains intriguing. Since jets from compact objects can accelerate particles to relativistic energies via Diffusive Shock Acceleration (DSA) mechanism~\citep{1978MNRAS.182..147B,1978ApJ...221L..29B}, we must ask whether MYSO jets can do the same, and whether these accelerated particles could ultimately produce $\gamma$-rays via processes such as relativistic Bremsstrahlung and Pion Decay. Indeed, several theoretical models proposing particle acceleration by MYSO jets and subsequent $\gamma$-ray production have been developed~\citep{2007A&A...476.1289A,2021MNRAS.504.2405A,2010A&A...511A...8B,2026arXiv260316647P}, and ~\citet{2026NatAs.tmp..133M} report a statistically significant detection of 
$\gamma$-rays from a population of young stellar objects. To date, HH 80-81 remains the only confirmed instance of $\gamma$-ray emission originating from an MYSO jet~\citep{2022RAA....22b5016Y,2025AA...695A..11M}. Additionally, two other MYSOs, S255 NIRS 3~\citep{2023MNRAS.523..105D} and AFGL 490~\citep{2026MNRAS.549ag896Y}, are considered highly likely to be associated with the $\gamma$-ray emission observed in their respective regions.

S140 is an HII region located at the southeastern edge of the L1204 molecular cloud, powered by the B0-type star HD~211880. Within this region, S140 IRS1 is an MYSO with a mass of $11 M_{\odot}$ and a bolometric luminosity of $8500 L_{\odot}$~\citep{2013MNRAS.428..609M}. It is located at a distance of $764 \pm 27$ pc~\citep{2008PASJ...60..961H}, and Galactic coordinates (l, b) = ($106.80^{\circ}, 5.31^{\circ}$). As a typical Class 0/I protostar, the age of S140 IRS1 is estimated to be $10^4-10^5$ years. Observations of S140 IRS1 have revealed an equatorial ionized disk wind, along with a separate large-scale bipolar CO molecular outflow (PA = 160$^{\circ}$/340$^{\circ}$) that is oriented perpendicular to the wind structure~\citep{2002A&A...381..905W}. The lower limit for the dynamical age of this outflow is estimated to be 6000 years~\citep{2002A&A...383..540P}. This jet-driven outflow is believed to be primarily responsible for the cavity and bow-shock structures found in the vicinity of S140 IRS1~\citep{2002A&A...381..905W}. The detection of strong H$_2$ shock emission near the same source~\citep{2002A&A...383..540P} further indicates that this outflow is undergoing a violent interaction with its ambient medium.

In this paper, we analysed 17.75 years of \textit{Fermi}-LAT data to understand the origin of the $\gamma$-ray emission near the MYSO S140 IRS1. The paper is organized as follows. In Sect.~\ref{data}, we present the data processing procedure for the $\gamma$-ray observations. In Sect.~\ref{gas}, we study the gas distribution in this region. In Sect.~\ref{origin}, we test both leptonic and hadronic models to describe the $\gamma$-ray emission, which is followed by discussion as presented in Sect.~\ref{discussion}. Conclusions are summarized in Sect.~\ref{conclusion}.

\begin{figure*}
    \centering
    \begin{tabular}{c@{\hspace{1em}}c@{\hspace{1em}}c}
        \multicolumn{1}{c}{(a)} & \multicolumn{1}{c}{(b)} & \multicolumn{1}{c}{(c)} \\[0.5ex]
        \includegraphics[width=0.28\textwidth]{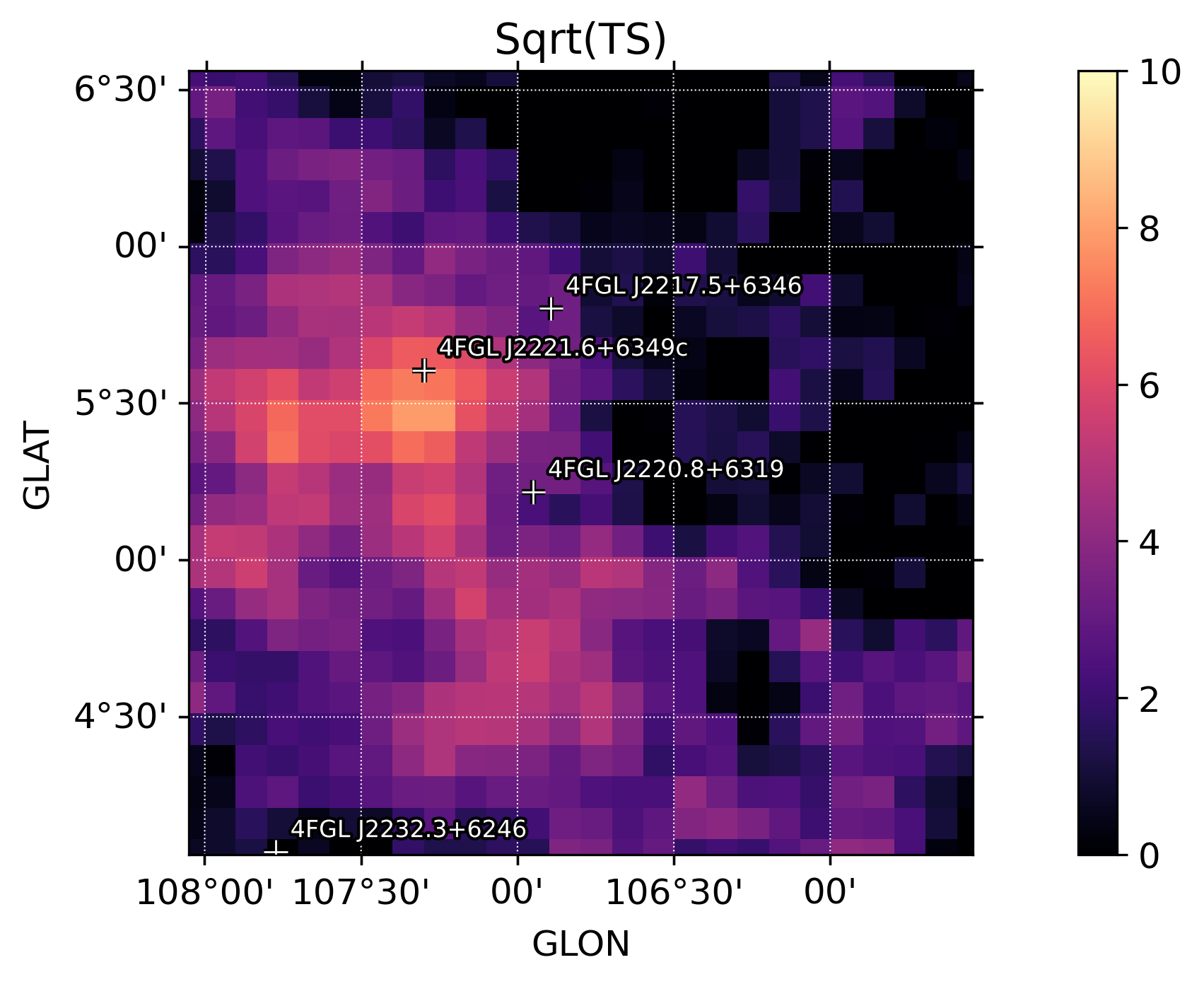} &
        \includegraphics[width=0.28\textwidth]{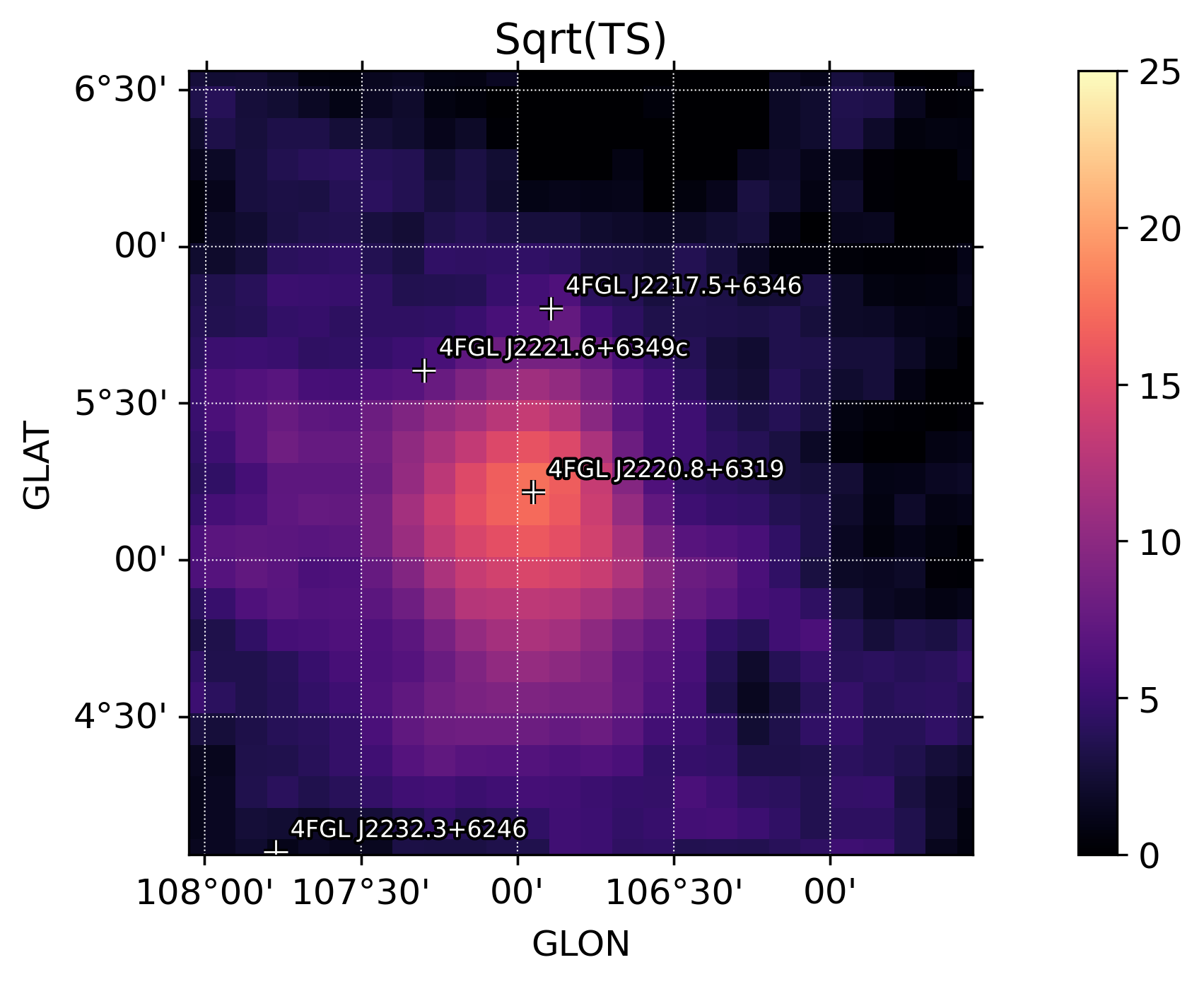} &
        \includegraphics[width=0.28\textwidth]{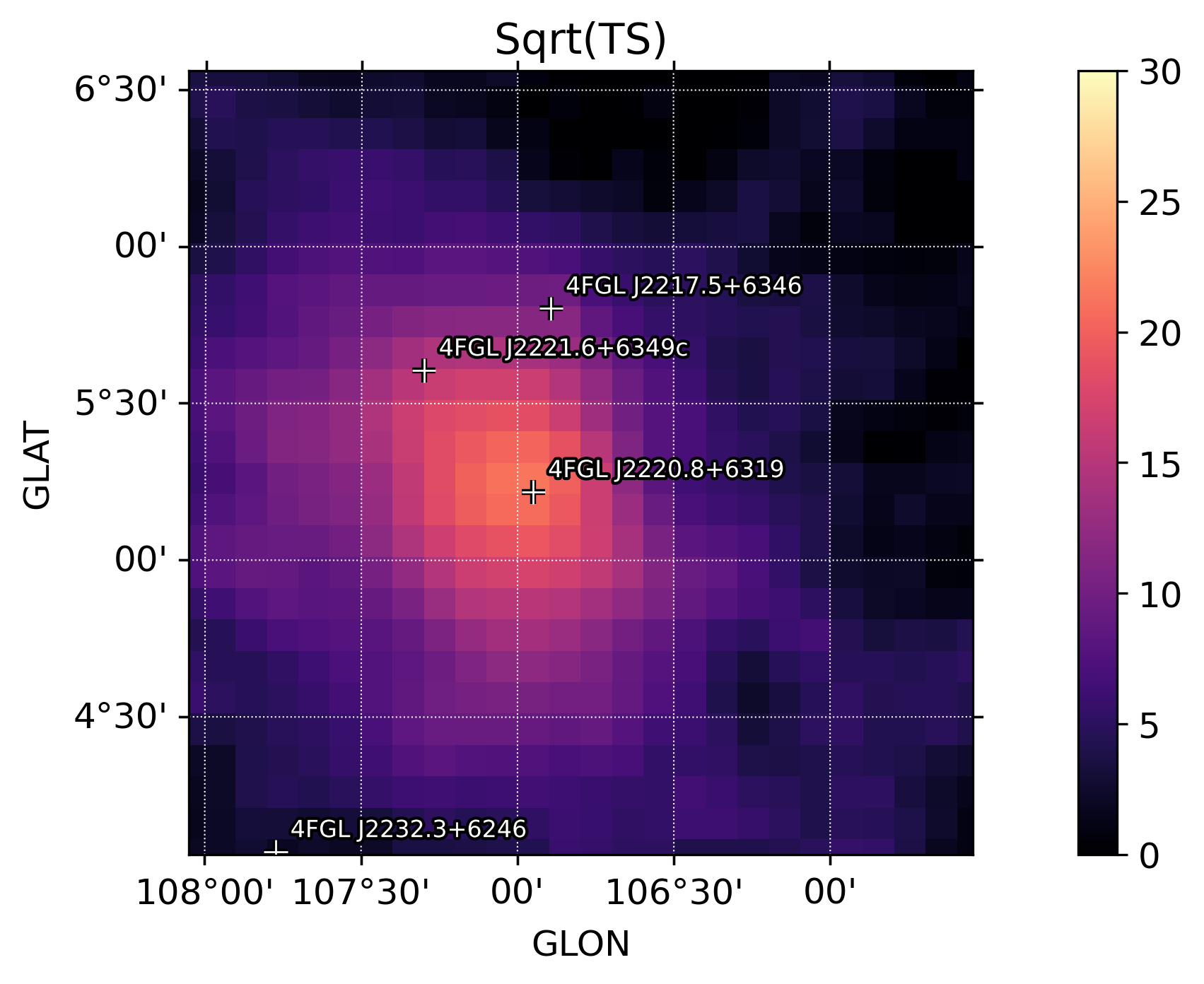} \\[1ex]
        \includegraphics[width=0.28\textwidth]{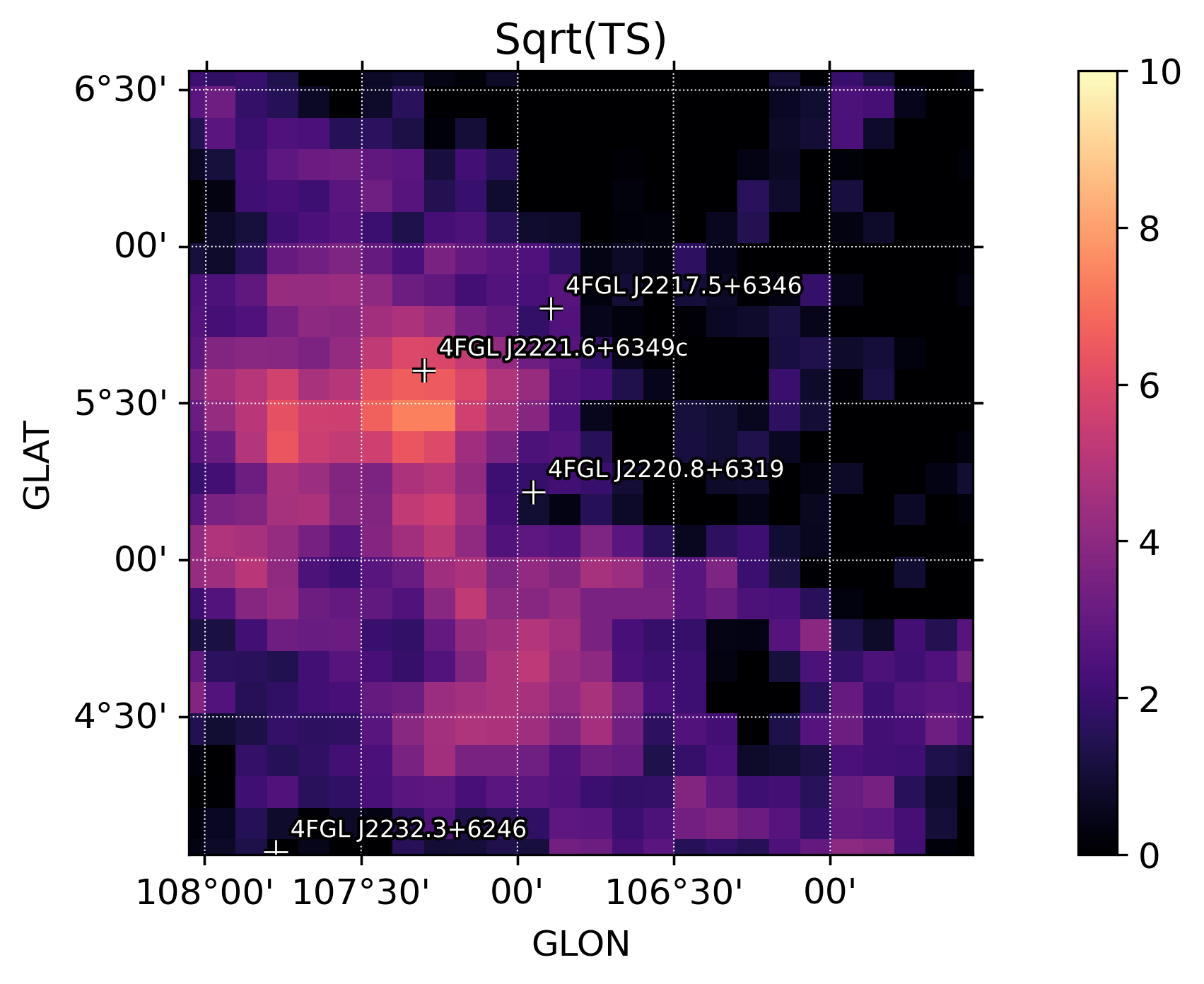} &
        \includegraphics[width=0.28\textwidth]{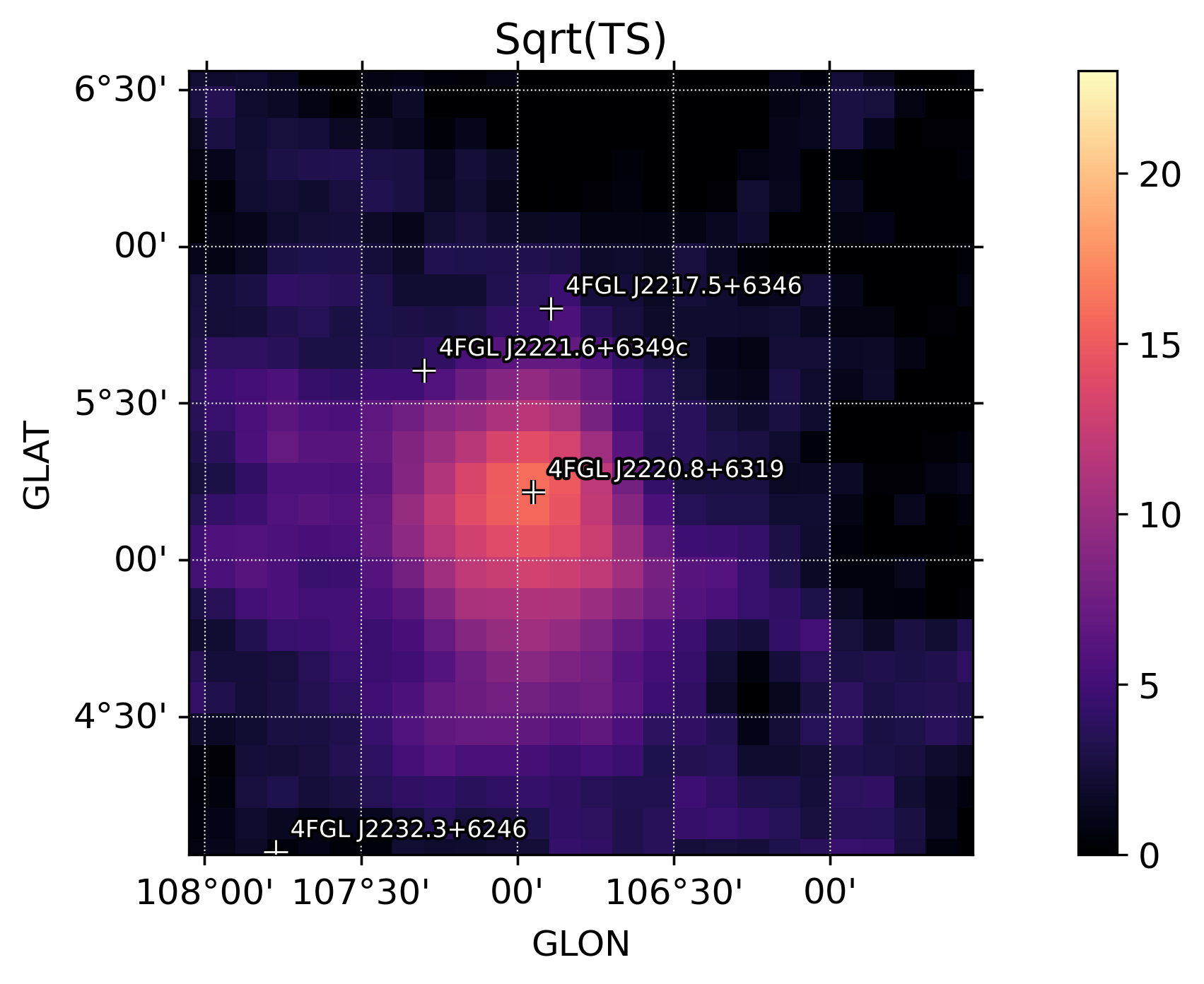} &
        \includegraphics[width=0.28\textwidth]{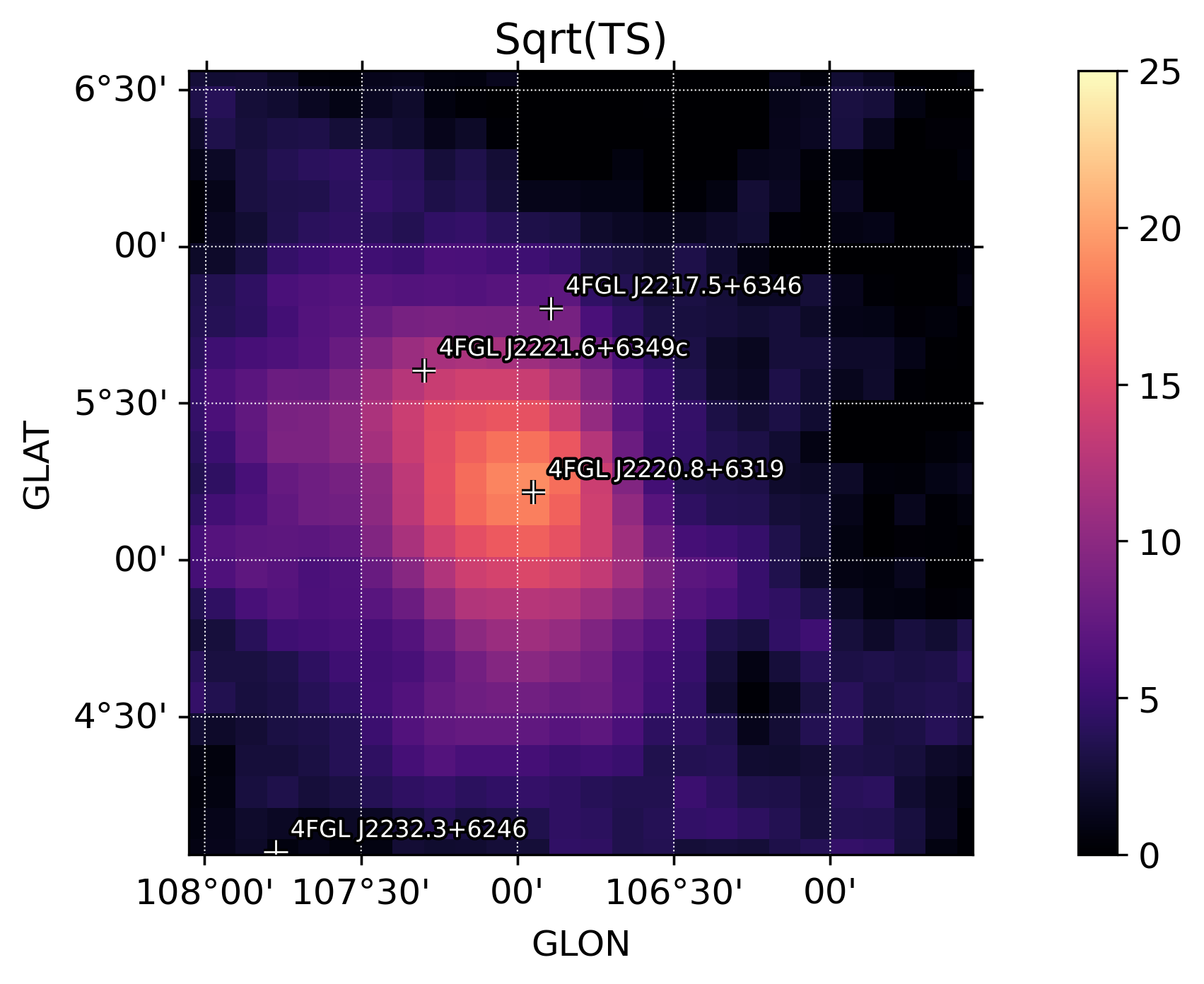} \\[1ex]
        \includegraphics[width=0.28\textwidth]{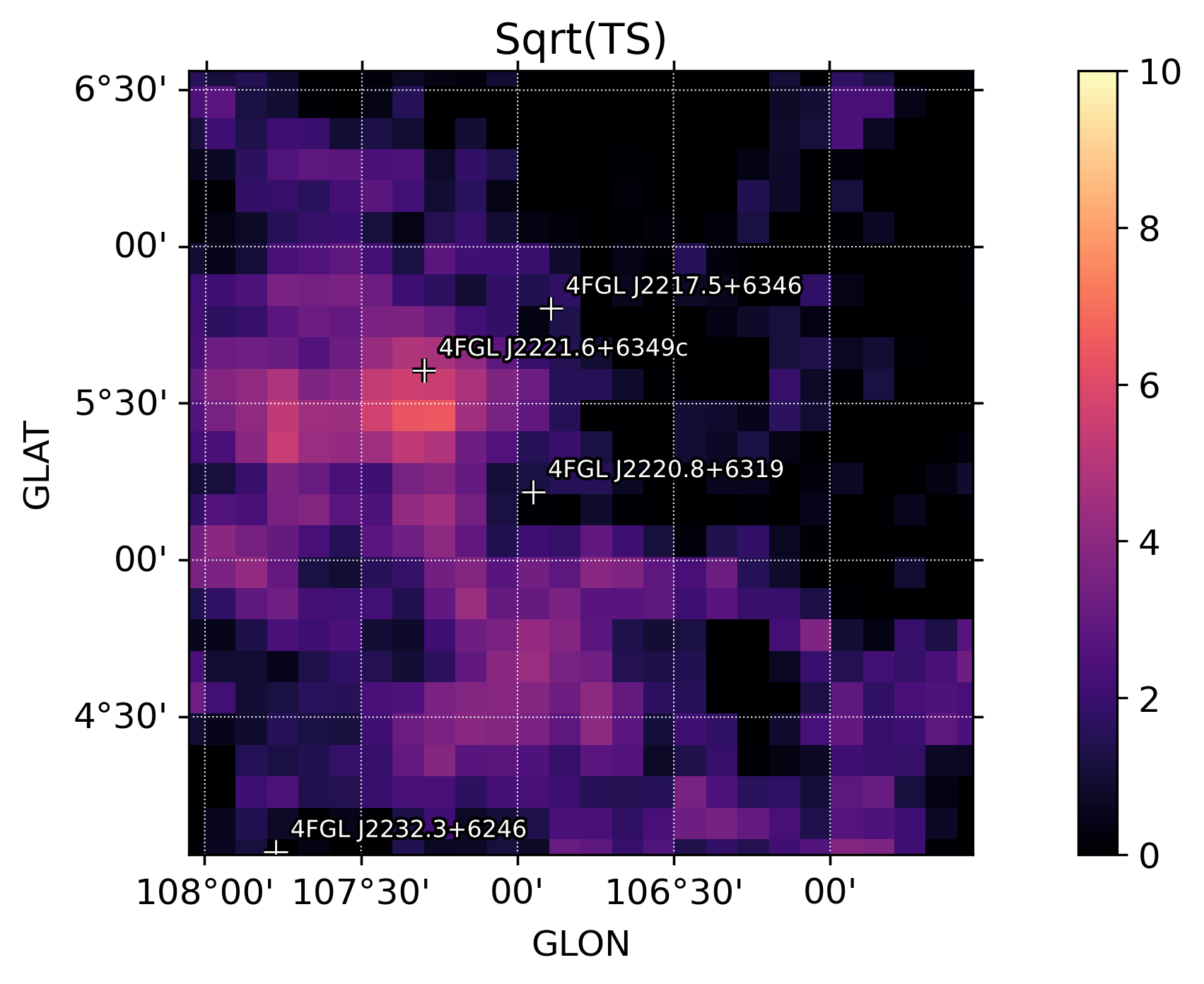} &
        \includegraphics[width=0.28\textwidth]{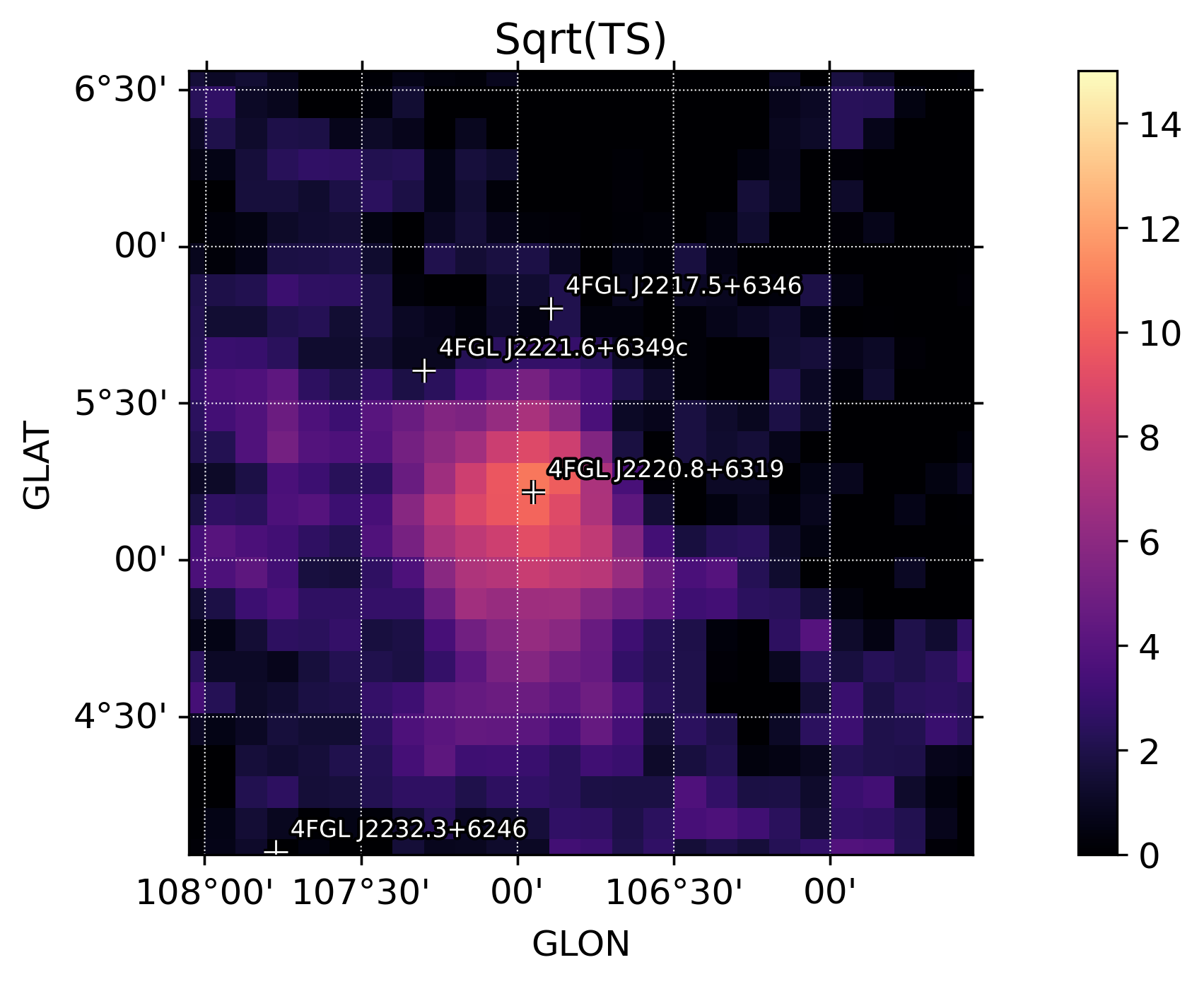} &
        \includegraphics[width=0.28\textwidth]{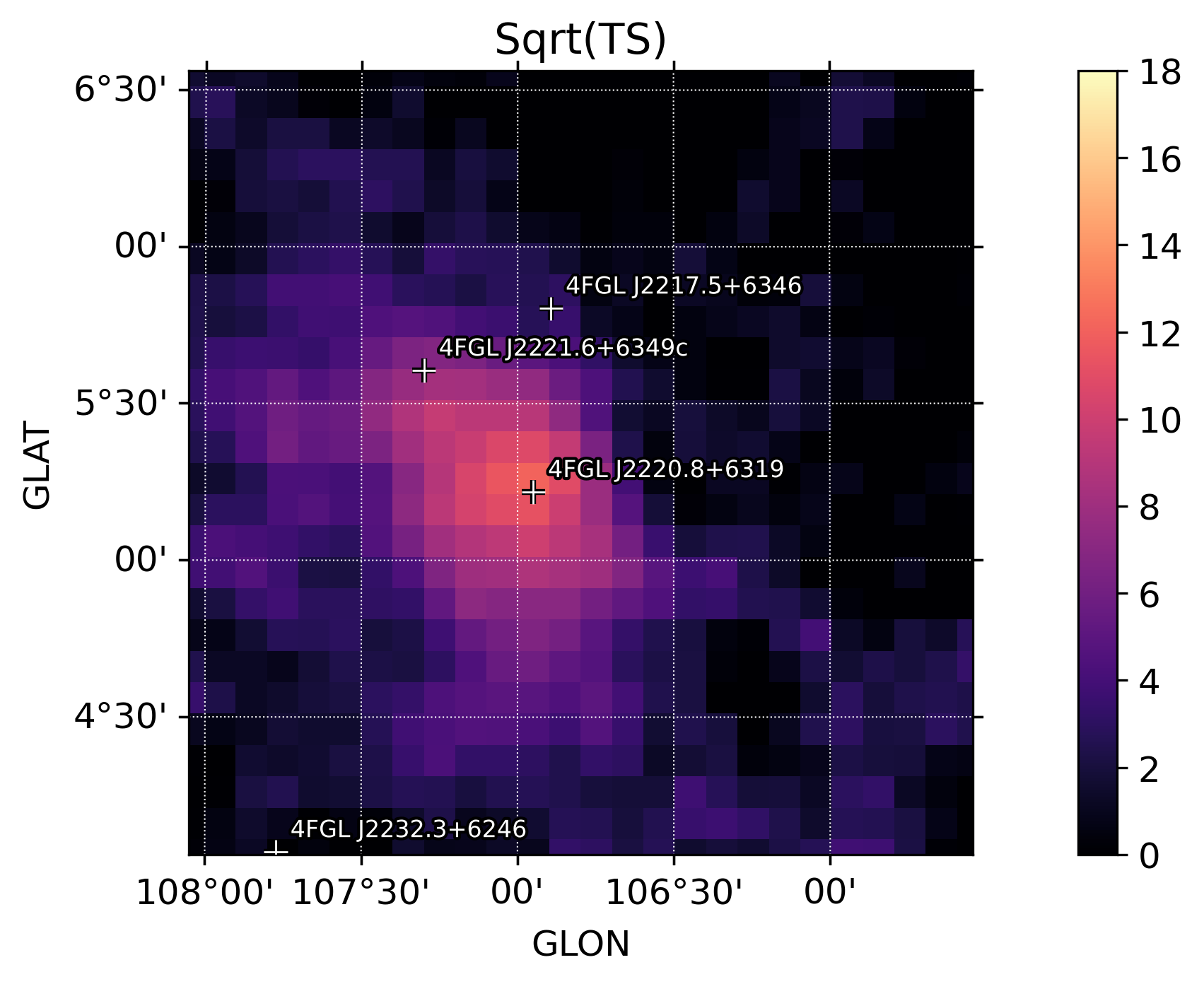}
    \end{tabular}
    \caption{$\sqrt{\mathrm{TS}}$ maps of the initial ROI ($2\degr \times 2\degr$) after the setup procedure. Columns (a), (b), and (c) correspond to the exclusion of 4FGL J2221.6+6349c only, 4FGL J2220.8+6319 only, and both 4FGL J2221.6+6349c and 4FGL J2220.8+6319, respectively. From top to bottom, each column shows the energy bands of 0.1--500\,GeV, 0.4--500\,GeV, and 1--500\,GeV. The test point sources are modeled as power-law spectra with a spectral index of 2.0.}
    \label{TS}
\end{figure*}

\section{Fermi-LAT DATA ANALYSIS}
\label{data}
The Fermi Large Area Telescope (\textit{Fermi}-LAT) is a space-borne \(\gamma\)-ray observatory that detects incident photons by measuring their arrival times, trajectories, and energies via its tracker and calorimeter subsystems, covering a broad spectrum from 20 MeV to over 1 TeV \citep{2009ApJ...697.1071A}. We use the Fermi Large Area Telescope Fourth Source Catalog (4FGL) as the reference catalog for source identification \citep{2020ApJS..247...33A}.  For this study, we selected the energy range of 0.1--500 GeV for our analysis. The data were downloaded through the Fermi Science Support Center (FSSC)\footnote{\url{https://fermi.gsfc.nasa.gov/ssc/data/}}. Our dataset  encompasses observations from 2008 August 6 (MET = 239673601) to 2026 April 6 (MET = 797126405), which we processed utilizing \textit{Fermipy} v1.4.0\footnote{\url{https://fermipy.readthedocs.io/en/latest/}}\citep{2017ICRC...35..824W}. The \texttt{P8R3\_SOURCE\_V3} instrument response functions (IRFs) were selected for the analysis of events and, to maximize the statistical significance, we chose events with \texttt{evtype = 3} and \texttt{evclass = 128}. Good Time Intervals were determined from the spacecraft file by applying the standard quality filter expression \texttt{(DATA\_QUAL > 0) \&\& (LAT\_CONFIG == 1)}. The entire analysis is confined to a \(16^{\circ}\times 16^{\circ}\) square Region of Interest (ROI) centered precisely on S140 IRS1.

\subsection{Preliminary analysis}
\label{preliminary}
In the subsequent analyses, we adopt the following fitting strategy: we allow the normalizations of all sources within a radius of $5^\circ$ from the ROI center (including the normalizations of the Galactic diffuse emission and the isotropic background components) to vary, and we also thaw all spectral parameters of sources within a radius of $1^\circ$ from the ROI center. All these free parameters are then fitted simultaneously. For clarity of presentation, we show a square region of $2^\circ \times 2^\circ$ centered on the ROI. The nearest $\gamma$-ray source to the ROI center (i.e., the MYSO S140 IRS1) is 4FGL J2220.8+6319, with Galactic coordinates $(l = 106.95^\circ, b = 5.22^\circ)$. Its angular separation from S140 IRS1 is $0.18^\circ$. 

Within this $2^\circ \times 2^\circ$ region, we focus on three $\gamma$-ray sources, whose spatial positions are shown in Fig.~\ref{TS}. We first perform a fit in the $0.1$--$500~\mathrm{GeV}$ energy band. The fitting results are as follows:
\begin{itemize}
    \item \textbf{4FGL~J2220.8+6319}: This is our target source, with $\mathrm{TS}=651.5$, $N_{\mathrm{pred}}=11909.5$, and a Log-Parabola(LP) spectrum:
    \begin{equation}\label{LP}
        \frac{dN}{dE} = N_0 \left( \frac{E}{E_b} \right)^{-\alpha - \beta \ln(E / E_b)} .
    \end{equation}
    The spectral parameters are $\alpha = 2.85$, $\beta = 0.40$, $E_b = 1385\,\mathrm{MeV}$;
    \item \textbf{4FGL~J2221.6+6349c}: $\mathrm{TS}=83.5$, $N_{\mathrm{pred}}=2601.0$, also with a LP spectrum, with parameters $\alpha = 2.38$, $\beta = 0.39$, $E_b = 1396\,\mathrm{MeV}$;
    \item \textbf{4FGL~J2217.5+6346}: $\mathrm{TS}=212.7$, $N_{\mathrm{pred}}=370.0$, also a LP spectrum, with $\alpha = 1.51$, $\beta = 0.42$, $E_b = 5269\,\mathrm{MeV}$. In addition, this source is marked as being associated with the X-ray source 1RXS~J221728.9+634714~\citep{2000IAUC.7432....3V}.
\end{itemize}

Next, we selected events with energies above 1 GeV and re-localized the above three sources using the \texttt{localize} method in \textit{Fermipy}. The positional offsets for 4FGL~J2220.8+6319 and 4FGL~J2217.5+6346 are $0.01^\circ$ and $0.02^\circ$, respectively, while that for 4FGL~J2221.6+6349c is $0.13^\circ$. This relatively large offset can be understood from Fig.~\ref{TS}, as its TS distribution appears diffuse. In the subsequent analysis, we retain the original positions of all three sources. On the other hand, by analyzing the above information, we can conclude that source 4FGL~J2221.6+6349c may be related to our target source 4FGL~J2220.8+6319, owing to the similarity in their spectral characteristics (in fact, there is also a spatial association, see Fig.~\ref{Hcontours} and Sec.~\ref{discussion}). On the other hand, source 4FGL~J2217.5+6346 is not considered to be associated with the target source 4FGL~J2220.8+6319, because of its significantly different spectral properties and the lack of apparent spatial correlation (see Fig.~\ref{Hcontours}).

\subsection{Comparison of different spatial models}
\label{Comparison}
First, we performed an extension test on the target source 4FGL J2220.8+6319. The \(TS_{\mathrm{ext}}\) values presented here are directly obtained from the output of the \texttt{Fermipy} tool. For the Gaussian extension (which corresponds to Model~2 in the subsequent comparison), the \(TS_{\mathrm{ext}}\) values in the \(>400\) MeV and \(>1\) GeV bands are 56.07 and 44.89, respectively; for the Disk extension (Model~3), the corresponding values are 48.33 and 37.20. Combined with Fig.~\ref{TS}, we tentatively conclude that the $\gamma$-ray emission in this region possesses an extended component, and we will consider different spatial extension templates to describe this extended emission component in the subsequent model comparison. Moreover, the spectra of both extended models exhibit significant curvature ($TS_{\mathrm{cur}}$ of the LP model exceeds 100 in each case, using the \texttt{curvature} method in \textit{Fermipy}). Hence, we fix the spectral shape of the target source to a log-parabola in the following model comparison. Finally, from Fig.~\ref{TS} we also note that it is evident that excluding only 4FGL J2221.6+6349c yields a poor result, appearing as a diffuse $\sqrt{\mathrm{TS}}$ distribution. Therefore, we also consider a model where 4FGL J2221.6+6349c is removed.

\begin{figure}
    \centering
    \includegraphics[width=0.5\textwidth]{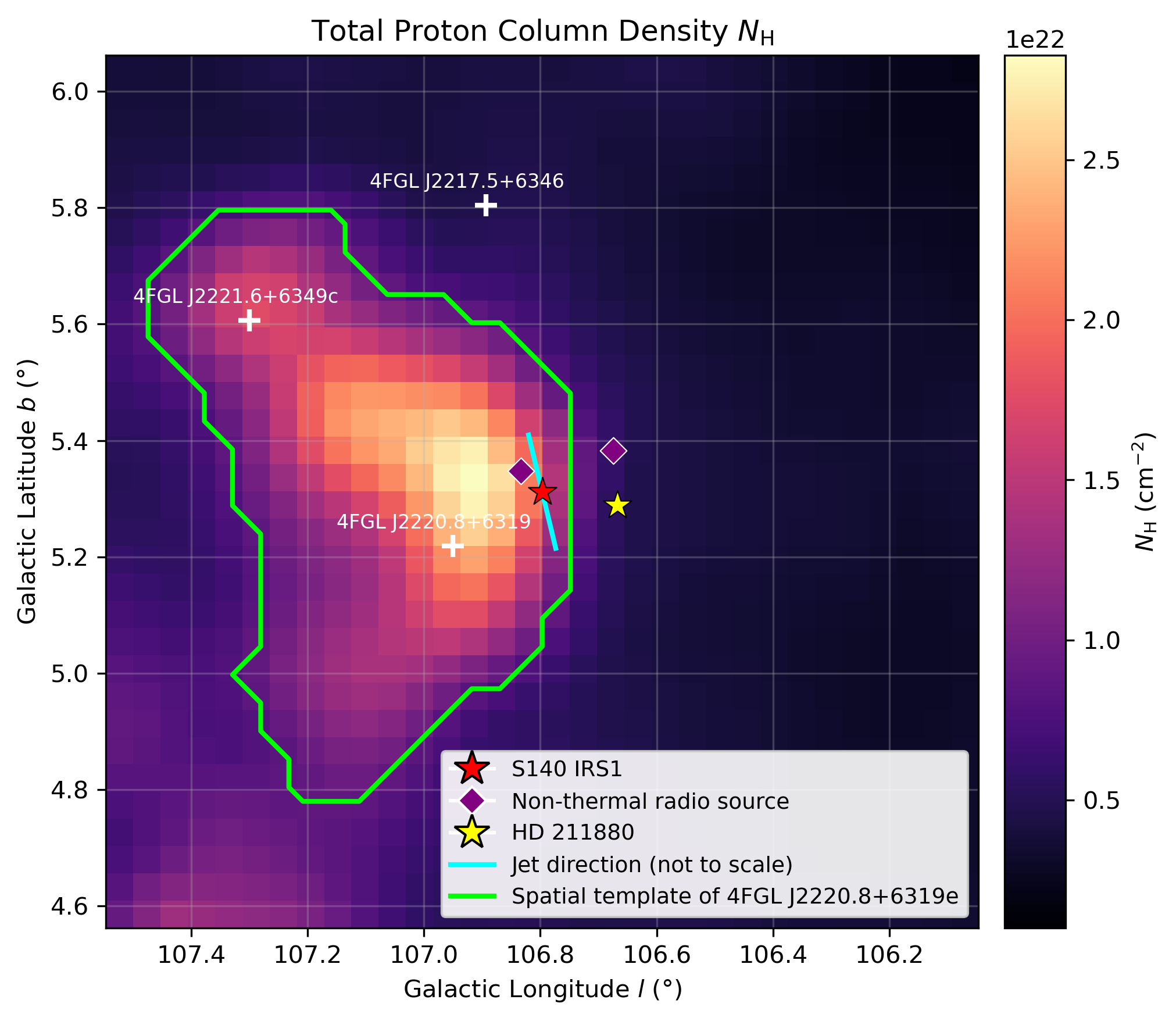}
    \caption{Map of the total proton column density within the central $1.5^\circ \times 1.5^\circ$ region centered on the ROI. For ease of comparison, the initial positions of the three point sources are also indicated with crosses. The star closer to the central cross marks the location of the MYSO S140~IRS1, and a solid line at this star position indicates the jet direction (not representing the actual size). The two diamonds on either side of the solid line correspond to two non-thermal radio point sources. The star to the right of the solid line marks the position of HD~211880. The irregular polygon outlines the boundary of the total proton template (see Sec.~\ref{Comparison} and Sec.~\ref{density}), which corresponds to the boundary of the spatial distribution model for 4FGL J2220.8+6319e.} 
    \label{Hcontours}
\end{figure}

We then employed the Akaike Information Criterion~\citep[AIC][]{1974ITAC...19..716A} to compare the following six models and determine which one most accurately describes the $\gamma$-ray emission in this region:
\begin{itemize}
    \item \textbf{Model 1}: the initial baseline model obtained after the first fit, which includes 4FGL J2220.8+6319 and 4FGL J2221.6+6349c, both treated as point sources;
    \item \textbf{Model 2}: based on Model 1, but with the spatial template of 4FGL J2220.8+6319 changed to a Gaussian extension;
    \item \textbf{Model 3}: based on Model 1, but with the spatial template of 4FGL J2220.8+6319 changed to a Disk extension;
    \item \textbf{Model 4}: based on Model 1, but with 4FGL J2221.6+6349c removed and the spatial template of 4FGL J2220.8+6319 changed to a Gaussian extension;
    \item \textbf{Model 5}: based on Model 1, but with 4FGL J2221.6+6349c removed and the spatial template of 4FGL J2220.8+6319 changed to a Disk extension;
    \item \textbf{Model 6}: based on Model 1, but with both 4FGL J2220.8+6319 and 4FGL J2221.6+6349c removed, and fitted using the spatial distribution template of total protons (i.e., \ce{H2}+\ce{HI}+\ce{HII}, see Sec.~\ref{density} and Fig.~\ref{Hcontours}), while still maintaining a log-parabola spectrum.
\end{itemize}

The obtained results are presented in Tab.~\ref{AIC}. To minimise the influence of the PSF and Galactic diffuse emission as much as possible, we primarily focus on data above 400 MeV and use these as the basis for model selection. Ultimately, we adopt the total proton template (Model~6) as our final template and perform spectral fitting over the full energy band (\(>100\) MeV), for the following reasons:
\begin{enumerate}
    \item Above 400 MeV, the AIC evidence ratio (ER) of Model~6 is of the same order of magnitude as that of Model~2, and is at least one order of magnitude larger than those of all the other models, making it the second-best model;
    \item Above 1 GeV, the ER of Model~6 is comparable to that of Model~2 and significantly exceeds those of all other models by several orders of magnitude;
    \item Overall, statistically speaking, Model~6 is comparable to Model~2, but the total-proton template (Model~6) can still be adopted because it has a clearer physical interpretation.
\end{enumerate}

\renewcommand{\arraystretch}{1.2}

\begin{table*}[t]
\centering
\caption{Comparison of different spatial models across three energy bands. The AIC is defined as \(\mathrm{AIC} = 2k - 2\ln L\), and the difference is calculated relative to Model~1 as \(\Delta\mathrm{AIC} = \mathrm{AIC}_{\mathrm{Model\,2-6}} - \mathrm{AIC}_{\mathrm{Model\,1}}\). Note that the total number of free parameters \(k\) consists of the parameters for the target sources (\(k_{\rm src}\)) and the background components (\(k_{\rm bg}\)). Since the background model parameters \(k_{\rm bg}\) remain unchanged across all fits (see Sec.~\ref{preliminary}), they introduce the same constant offset to the AIC values and cancel out in the calculation of \(\Delta\mathrm{AIC}\). Thus, only the source parameters \(k_{\rm src}\) are listed in the table. Specifically, we fix the positions of all sources, so the listed \(k\) values solely account for the spectral parameters (3 for each LogParabola source) and the spatial morphology parameters (1 for a fitted Gaussian or disk extension, and 0 for a fixed spatial template). The values of \(\ln L\) are directly obtained from the \texttt{Fermipy} output. To quantify the relative support for each model, we compute the AIC evidence ratio (ER) as \(\mathrm{ER} = \exp(-\Delta\mathrm{AIC}/2)\), which represents the relative likelihood of the alternative model compared to the baseline~\citep{2007MNRAS.377L..74L}. \label{AIC}}
\resizebox{\textwidth}{!}{%
\begin{tabular}{l c | ccc | ccc | ccc}
\hline
\multirow{2}{*}{\textbf{Configuration}} & \multirow{2}{*}{\(k_{\rm src}\)} & \multicolumn{3}{c|}{\(> 100\) MeV} & \multicolumn{3}{c|}{\(> 400\) MeV} & \multicolumn{3}{c|}{\(> 1\) GeV} \\
 & & \(\ln L\) & \(\Delta\)AIC & ER & \(\ln L\) & \(\Delta\)AIC & ER & \(\ln L\) & \(\Delta\)AIC & ER \\
\hline
Model 1 (Two point sources; Baseline)      & 6 & 17202424.151 & 0.00   & 1.00 & 5453028.069 & 0.00  & 1.00 & 895011.072 & 0.00   & 1.00 \\
Model 2 (Point source + Gaussian extension) & 7 & 17202452.681 & -55.06 & \(9.04\times 10^{11}\) & 5453055.865 & -53.59 & \(4.34\times 10^{11}\) & 895033.454 & -42.76 & \(1.93\times 10^{9}\) \\
Model 3 (Point source + Disk extension)     & 7 & 17202448.653 & -47.00 & \(1.61\times 10^{10}\) & 5453051.944 & -45.75 & \(8.60\times 10^{9}\) & 895029.542 & -34.94 & \(3.86\times 10^{7}\) \\
Model 4 (Gaussian extension only)           & 4 & 17202452.172 & -60.04 & \(1.09\times 10^{13}\) & 5453044.116 & -36.09 & \(6.88\times 10^{7}\) & 895026.846 & -35.55 & \(5.24\times 10^{7}\) \\
Model 5 (Disk extension only)               & 4 & 17202445.578 & -46.85 & \(1.49\times 10^{10}\) & 5453037.913 & -23.69 & \(1.39\times 10^{5}\) & 895018.512 & -18.88 & \(1.26\times 10^{4}\) \\
Model 6 (Total proton template)    & 3 & 17202444.112 & -45.92 & \(9.37\times 10^{9}\) & 5453050.689 & -51.24 & \(1.34\times 10^{11}\) & 895029.489 & -42.83 & \(2.00\times 10^{9}\) \\
\hline
\end{tabular}%
}
\end{table*}

\subsection{Spectral energy distribution}
\label{sec:spectral}
We designate the $\gamma$-ray source obtained from the total proton template fit as 4FGL~J2220.8+6319e, with a TS value of 1029.9. Its spectrum is described by a log-parabola, and our spectral fitting yielded the following parameter values: $N_0 = (2.38 \pm 0.10) \times 10^{-12}\ \mathrm{MeV^{-1}\ cm^{-2}\ s^{-1}}$, $\alpha = 2.44 \pm 0.05$, $\beta = 0.29 \pm 0.03$, and $E_b = 1200$~MeV. The spectral curvature test yields $TS_{\mathrm{cur}}(\mathrm{LP}) = 104.1$ and $TS_{\mathrm{cur}}(\mathrm{PLSC})$\footnote{\url{https://fermi.gsfc.nasa.gov/ssc/data/analysis/scitools/source_models.html}} $= 89.2$, indicating that significant curvature still remains. We will subsequently analyze 4FGL~J2220.8+6319e.

\section{GAS CONTENT}
\label{gas}
We investigated three distinct gas phases---molecular hydrogen (\ce{H2}), neutral atomic hydrogen (\ce{HI}), and ionized hydrogen (\ce{HII})---in the vicinity of the MYSO S140 IRS1.

\subsection{The \ce{H2} distribution}
We use the Planck CO revisited data \citep{2024A&A...688A..54G} with a $4\sigma$ mask to trace the molecular hydrogen distribution around S140 IRS1. The data are directly provided as CO (1$\to$0) integrated intensity $W_{\mathrm{CO}}$ (in K km s$^{-1}$), which we convert to molecular hydrogen column density using a conversion factor $X_{\mathrm{CO}} = 2\times 10^{20}~\mathrm{cm^{-2}~K^{-1}~km^{-1}~s}$ \citep{2013ARA&A..51..207B}. For comparison, we also generated a map from the Dame CO survey data \citep{2001ApJ...547..792D}. In that case, we extracted the CO spectrum toward the molecular cloud L1204 ($l=107.47^\circ$, $b=4.81^\circ$) and integrated over the velocity range $[-7.8, -3.9]~\mathrm{km~s^{-1}}$ where the signal-to-noise ratio exceeds $3\sigma$. After applying the same $X_{\mathrm{CO}}$ factor, the resulting column density distribution is similar to the one derived from the Planck data. Because the Planck CO revisited data offer superior quality in both resolution and signal-to-noise ratio, we ultimately adopt the Planck-based map for our subsequent analyses.

\subsection{The \ce{HI} and \ce{HII} distribution}
To trace the spatial distribution of neutral hydrogen, we use the HI4PI~\citep{2016A&A...594A.116H} 21\,cm data cube and adopt the following formula to compute the HI column density~\citep{1990ARA&A..28..215D}:
\begin{equation}
N_{\mathrm{HI}} = -\,1.823 \times 10^{18}\, T_s
\int \ln \left[ 1 - \frac{T_B}{T_s - T_{\mathrm{bg}}} \right] \, \mathrm{d}v ,
\end{equation}
where $T_{\mathrm{bg}} \approx 2.66\ \mathrm{K}$ is the 21\,cm CMB brightness temperature, and $T_B$ denotes the brightness temperature of the HI emission. When $T_B > T_s - 5\ \mathrm{K}$, it is truncated to this value, and we fix the spin temperature at $T_s = 150\ \mathrm{K}$. The integration is performed over the velocity range $[-12.78, 2.68]\ \mathrm{km\,s^{-1}}$ where the signal-to-noise ratio exceeds $3\sigma$.

To map the ionized hydrogen distribution, we use the Planck free-free map~\citep{2016A&A...594A..10P}. We first convert the emission measure (EM) into the free-free intensity using the conversion factor provided in Table~1 of \citet{2003ApJS..146..407F}. Then we derive the HII column density from the free-free intensity ($I_\nu$) following Eq.~(5) of \citet{1997ApJ...480..173S}:
\begin{equation} 
\begin{aligned} 
N_{\mathrm{H\,II}} &= 1.2 \times 10^{15}\ \mathrm{cm^{-2}}
\left( \frac{T_e}{1\ \mathrm{K}} \right)^{0.35}
\left( \frac{\nu}{1\ \mathrm{GHz}} \right)^{0.1} \\ 
&\quad \times \left( \frac{n_e}{1\ \mathrm{cm^{-3}}} \right)^{-1}
\frac{I_\nu}{1\ \mathrm{Jy\ sr^{-1}}} \ ,
\end{aligned}
\end{equation}
where $\nu = 353\ \mathrm{GHz}$ is the frequency, and the electron temperature is $T_e = 8000\ \mathrm{K}$. We adopt an effective electron density of $2\ \mathrm{cm^{-3}}$, as suggested by \citet{1997ApJ...480..173S} for the region outside the solar circle.

With the above procedures, we obtain the spatial distribution maps of the neutral hydrogen and ionized hydrogen column densities. We find that both distributions exhibit a diffuse background and show no significant correlation with either the Sqrt(TS) map (see Fig.~\ref{TS}) or the molecular hydrogen column density map.

\subsection{Estimating total proton number density} 
\label{density}
We construct the total proton column density map by co‑aligning the three hydrogen components (H\,{\sc i}, H\,{\sc ii}, and H$_2$) onto a common $32\times32$ pixel grid with a sampling of $0.048^\circ$. The total proton column density is then obtained as $N_{\rm H} = N_{\rm HI} + 2N_{\rm H_2} + N_{\rm HII}$, and its spatial distribution is displayed in Fig.~\ref{Hcontours}. To isolate the dense gas structure associated with the target source and suppress diffuse background and noise, we estimate the noise level of the $N_{\rm H}$ map using the median absolute deviation (MAD) and apply a $4\sigma_{\rm MAD}$ clipping threshold ($\approx 9\times10^{21}~\mathrm{cm^{-2}}$). The resulting compact structure defines the total proton template adopted in Sec.~\ref{Comparison}, and its boundary is overlaid on Fig.~\ref{Hcontours}.

Based on a distance of $d = 764~\mathrm{pc}$, we estimate the equivalent radius of this structure as $R = \sqrt{A/\pi}$, where $A$ is its projected area, yielding $R \sim 5.15~\mathrm{pc}$. Assuming a uniform spherical geometry ($V = \frac{4}{3}\pi R^{3}$) and a total proton mass $M_{\rm P} = M_{\rm H_2} + M_{\rm HI} + M_{\rm HII}$, we derive an average proton number density of $n_{\rm p} \sim 720~\mathrm{cm^{-3}}$.

\section{THE ORIGIN OF $\gamma$-RAY EMISSION}
\label{origin}
We use the \textsc{Naima} package \citep{2015ICRC...34..922Z} to fit the spectral energy distribution (SED) of the source 4FGL J2220.8+6319e with both leptonic and hadronic processes. For the leptonic scenario, given the lack of massive OB stars in the region and the consequently weak UV radiation field, the efficiency of the Inverse Compton scattering process is expected to be low. We therefore consider only the relativistic Bremsstrahlung process. For the hadronic scenario, we model the emission via proton-proton interactions (Pion Decay). We adopt the total proton number density derived in Sect.~\ref{density} as the target particle density for both processes. For both scenarios, we test five different particle distribution models and evaluate the goodness of fit using the Bayesian Information Criterion (BIC)~\citep{1978AnSta...6..461S}. The results are presented in Tab.~\ref{diffrent model}.

Based on the results in Tab.~\ref{diffrent model}, we adopt the Log-Parabolic (LP) spectral form to describe the electron distribution:
\begin{equation}
\label{LogParabola}
f(E) = A \left( \frac{E}{E_0} \right)^{-\alpha - \beta \ln\left( \frac{E}{E_0} \right)},
\end{equation}
where $E_0 = 1.4\ \text{GeV}$, the electron spectral index $\alpha = 1.98_{-0.08}^{+0.07}$, and $\beta = 0.45_{-0.06}^{+0.07}$. The total electron energy above $1\ \text{GeV}$ is $W_e = 1.04_{-0.04}^{+0.04} \times 10^{45}\ \text{erg}$, and the corresponding maximum log-likelihood value for this fit is $-3.80$.

For the proton distribution, we employ the Exponential Cutoff Power-Law (ECPL) form:
\begin{equation}
\label{ECPL}
f(E) = A \left( \frac{E}{E_0} \right)^{-\alpha} \exp\left( -\frac{E}{E_{\text{cutoff}}} \right),
\end{equation}
with $E_0 = 1.4\ \text{GeV}$, the proton spectral index $\alpha = 2.39_{-0.15}^{+0.14}$, and the cutoff energy $E_{\text{cutoff}} = 34.93_{-9.01}^{+13.83}\ \text{GeV}$. The total proton energy above $1\ \text{GeV}$ is $W_p = 2.29_{-0.18}^{+0.18} \times 10^{46}\ \text{erg}$, and the corresponding maximum log-likelihood value for this fit is $-1.77$.

\begin{figure}
    \centering
    \includegraphics[width=0.5\textwidth]{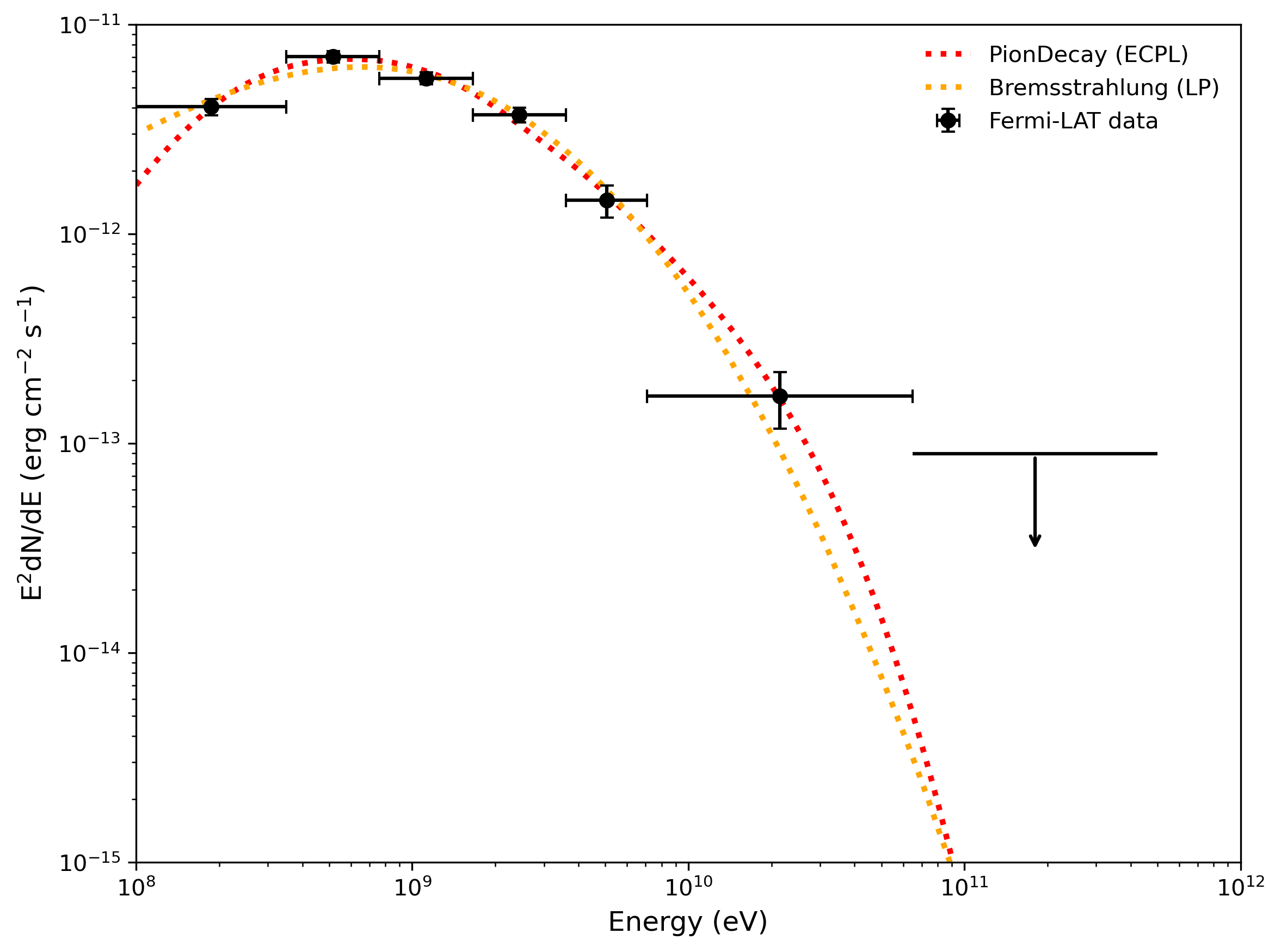}
    \caption{The spectral energy distribution (SED) resulting from the Pion Decay and relativistic Bremsstrahlung of 4FGL J2220.8+6319e.}
    \label{brems+pp}
\end{figure}

The fitted SED results for both processes are shown in Fig.~\ref{brems+pp}. It should be emphasized that the quoted uncertainties on the total particle energies do not include the systematic uncertainty propagated from the gas density (\(n_p\)) estimated in Sect.~\ref{density}. As a result, the derived total particle energies, \(W_e\) and \(W_p\), can only be regarded as order-of-magnitude estimates. Furthermore, given that the spectral fitting is performed with only six SED data points, the BIC alone is insufficient to robustly distinguish between the hadronic and leptonic scenarios, and our interpretation therefore remains qualitative at this stage.

\begin{table}
\centering
\caption{BIC values for different models.}
\label{diffrent model}
\begin{tabular}{lccccc}
\hline
         & PL & LP & ECPL & BPL & ECBPL\\
\hline
Bremsstrahlung & 263.47 & 13.44 & 25.08 & 83.43 & 85.15 \\
PionDecay & 45.28 & 11.38 & 9.38 & 13.48 & 13.02 \\
\hline
\end{tabular}
\end{table}

\section{DISCUSSION}
\label{discussion}

\subsection{Stellar winds VS Jet}
We first searched the Green's Galactic Supernova Remnant Catalog~\citep{2025JApA...46...14G} and the ATNF Pulsar Catalog~\citep{2005AJ....129.1993M} within the extension radius of 4FGL~J2220.8+6319e. No plausible $\gamma$-ray counterparts were found. Furthermore, we can argue against a significant influence of 4FGL~J2217.5+6346 on 4FGL~J2220.8+6319e based on the following considerations: (1) as shown in Fig.~\ref{Hcontours}, 4FGL~J2217.5+6346 does not appear to be spatially associated with the dense molecular gas; (2) the spectral properties of 4FGL~J2217.5+6346 are notably different from those of 4FGL~J2220.8+6319e. Based on these considerations, we turn our attention to possible particle accelerators within the extended source:

\begin{itemize}
    \item \textbf{Stellar winds from the S140 cluster:} S140 is a low- to intermediate-mass embedded cluster, whose most massive member is the MYSO S140~IRS1 with a mass of $11\,M_\odot$, while the other main members, IRS2--IRS7, have an average mass of $\sim5\,M_\odot$~\citep{2002A&A...383..540P}; there are also lower-mass members, but they are difficult to resolve and characterize. Since detailed physical properties are available only for IRS1, we estimate its wind power and reasonably assume that the total wind power of the cluster is of the same order of magnitude. Adopting the spherical wind model from \citet{2007MNRAS.380..246G}, we calculate the wind power of S140~IRS1 as
    \begin{equation}
        L_w = \frac{1}{2}\dot{M}_w v_w^2,
    \end{equation}
    where $\dot{M}_w = 3\times10^{-6}\,M_\odot\,\mathrm{yr}^{-1}$ is the mass-loss rate of the wind and $v_w = 400\,\mathrm{km\,s}^{-1}$ is the wind terminal velocity, yielding $L_w = 1.5\times10^{35}\,\mathrm{erg\,s}^{-1}$. We thus estimate the total stellar wind power of the S140 cluster as $L_{w,\mathrm{cluster}} \sim 2\times10^{35}\,\mathrm{erg\,s}^{-1}$. Taking the cluster age $t_{\mathrm{cluster}}=10^5\,\mathrm{yr}$ and the total proton energy $W_p$ obtained in Sec.~\ref{origin}, the proton acceleration efficiency can be derived as $\eta_{\mathrm{wind}} = W_p/(L_{w,\mathrm{cluster}}\,t_{\mathrm{cluster}}) \sim 3.6\%$. Such an efficiency is unreasonably high for a cluster that does not even contain a protostar of O-type mass, because typical Young Massive Star Clusters suggest that the efficiencies of particle acceleration by collective stellar winds are generally below $10\%$, and in some cases even lower than $1\%$~\citep{2020SSRv..216...42B}; there is currently no reason to believe that the S140 cluster can achieve such a level of acceleration efficiency.

    \item \textbf{The jet from S140~IRS1:} Although other protostellar jets may exist in the S140 region, the jet power of S140~IRS1 dominates in magnitude. We first compute the accretion luminosity of S140~IRS1 using
    \begin{equation}
        L_{\mathrm{acc}} = \frac{G M_\star \dot{M}_{\mathrm{acc}}}{R_\star},
    \end{equation}
    where $\dot{M}_{\mathrm{acc}}=7.5\times10^{-5}\,M_\odot\,\mathrm{yr}^{-1}$ is the disk accretion rate, $M_\star=11\,M_\odot$ is the mass of S140~IRS1, and $R_\star=4.92\,R_\odot$ is its radius, all adopted from \citet{2013MNRAS.428..609M}. This gives $L_{\mathrm{acc}} \sim 5300\,L_\odot$. Assuming a ratio of 0.075 between the jet kinetic luminosity and the accretion luminosity \citep{2007LNP...723...21C}, the kinetic luminosity of the MYSO S140~IRS1 jet is estimated to be $L_j = 1.5\times10^{36}\,\mathrm{erg\,s}^{-1}$. Adopting a jet acceleration efficiency in the range $2\%$--$10\%$ \citep{2021MNRAS.504.2405A} and using $t_{\mathrm{injection}} = W_p/(\eta L_{\mathrm{j}})$, we derive $t_{\mathrm{injection}} = 4.8\times10^3$--$2.4\times10^4$~yr, which is consistent with the age of MYSO~S140~IRS1. We further evaluate the maximum energy attainable by particles accelerated in the MYSO jet of S140~IRS1 using the criterion \citep{2025ApJ...989L..25W}
\begin{equation}
L_{\mathrm{j}} \geq 10^{38} \left( \frac{E_{\mathrm{max}}}{10\ \mathrm{PeV}} \right)^2 \widetilde{\omega} \beta^{-1} \sigma_{-1}^{-1} \ \mathrm{erg\ s^{-1}},
\end{equation}
where $\widetilde{\omega}=1$ is adopted for a collimated jet, $\beta \approx 0.0033$ (i.e., $v = 1000\ \mathrm{km/s}$), and $\sigma_{-1} = \sigma/0.1$ with the non-relativistic magnetization $\sigma = B^2/(2\pi\rho v^2)$. Assuming a mean magnetic field $B = 0.1\ \mathrm{mG}$ \citep{2019MNRAS.482.4687R} and a jet particle density $n_{\mathrm{j}} = 10^{4}\ \mathrm{cm^{-3}}$ \citep{2021MNRAS.504.2405A}, the kinetic luminosity $L_{\mathrm{j}} = 1.5\times10^{36}\,\mathrm{erg\,s^{-1}}$ gives $E_{\mathrm{max}} \approx 0.55\ \mathrm{TeV}$, in agreement with the typical $\sim$0.24~TeV proton maximum energy expected for protostellar jet acceleration \citep{2021MNRAS.504.2405A}. For our ECPL proton and LP electron models, we define the maximum energy as the value containing 99.99\% of the total distribution, yielding $E_{p,\mathrm{max}} = 85.76\ \mathrm{GeV}$ and $E_{e,\mathrm{max}} = 127.73\ \mathrm{GeV}$, both comfortably within the derived limit.
\end{itemize}

As for the jet scenario, we also searched for possible non-thermal radio sources within a $10'$ radius centered on MYSO~S140~IRS1, since their presence would imply particles accelerated to relativistic energies. The results are as follows:
\begin{itemize}
    \item At a distance of $\sim 0.05^\circ$ from the MYSO, we identified the radio source NVSS~J221924+632143~\citep{1998AJ....115.1693C}, with an integrated total flux of $58.4\,\mathrm{mJy}$ in the $L$ band ($1.4\,\mathrm{GHz}$). Coincident with this position (the only counterpart within $2'$, separated by $\sim 1''$) is VLASS1QLCIR~J221924.20+632144.8~\citep{2021ApJS..255...30G}, which has an integrated total flux of $34.196\,\mathrm{mJy}$ in the $S$ band ($2$--$4\,\mathrm{GHz}$), along with a unique $74\,\mathrm{MHz}$ VLSSr source~\citep{2014MNRAS.440..327L} whose total flux, estimated from the peak intensity, is $0.64\,\mathrm{Jy}$. Assuming all three represent the same radio source, we derived spectral indices $\alpha_{74-L} = -0.81\pm 0.05$ and $\alpha_{L-S} = -0.70\pm 0.06$.
    \item At a distance of $\sim 0.14^\circ$ from the MYSO, we found NVSS~J221803+631814~\citep{1998AJ....115.1693C}, with an integrated total flux of $48.8\,\mathrm{mJy}$ in the $L$ band ($1.4\,\mathrm{GHz}$). Its positional coincidence (the only counterpart within $2'$, separated by $\sim 1''$) is VLASS1QLCIR~J221803.48+631815.0~\citep{2021ApJS..255...30G}, with an integrated total flux of $28.415\,\mathrm{mJy}$ in the $S$ band ($2$--$4\,\mathrm{GHz}$). No $74\,\mathrm{MHz}$ data exist within $2'$. We derived a spectral index $\alpha_{L-S} = -0.71\pm 0.04$.
\end{itemize}
Judging from their spectral indices, both radio sources display non-thermal characteristics. Their locations are indicated in Fig.~\ref{Hcontours}. As seen in the figure, the non-thermal source closer to the MYSO lies near the jet direction and could be produced by particles accelerated in the jet, whereas the other non-thermal source is situated farther from the jet.

\subsection{Diffusion of high-energy particles}
we examine the diffusion of high-energy particles within the molecular cloud. The diffusion radii of protons and electrons are computed respectively as~\citet{2004vhec.book.....A}:
\begin{equation}
\label{pdiff}
R_{\mathrm{diff,p}}(E, t)
= 2 \sqrt{ D(E)\, t \,
\frac{ e^{t\delta/t_{\mathrm{pp}}} - 1 }{ t\delta / t_{\mathrm{pp}} } } ,
\end{equation}
\begin{equation}
\label{ediff}
R_{\mathrm{diff,e}}(E, t)
\simeq 2 \sqrt{ D(E)\, t \,
\frac{ 1 - \left( 1 - E/E_{\mathrm{cut}} \right)^{1-\delta} }
{ (1-\delta)\, E/E_{\mathrm{cut}} } } .
\end{equation}
Here the diffusion coefficient is $D(E)=D_{0}\,(E/10~\mathrm{GeV})^{\delta}$, with $D_{0}$ typically in the range $10^{26}$--$10^{28}\,\mathrm{cm^{2}\,s^{-1}}$. To obtain a conservative lower limit, we adopt $D_{0}=10^{26}\,\mathrm{cm^{2}\,s^{-1}}$ (appropriate for a dense environment) and $\delta=0.5$, and consider a diffusion time $t = 10^{4}$--$10^{5}\,\mathrm{yr}$.

For protons, the $pp$ energy-loss timescale is $t_{\mathrm{pp}} = (n_{0}\,\sigma_{\mathrm{pp}}\,f\,c)^{-1} \simeq 5.3\times10^{7}\,(n/1\,\mathrm{cm^{-3}})^{-1}\,\mathrm{yr}$. Using the density $n \sim 720\,\mathrm{cm^{-3}}$ derived in Sec.~\ref{density} yields $t_{\mathrm{pp}} \approx 7.4\times10^{4}\,\mathrm{yr}$. For electrons, the time-dependent cutoff energy is $E_{\mathrm{cut}}(t) \simeq m_{\mathrm{e}}c^{2}/[(7\times10^{-20}\,\mathrm{eV\,cm^{-3}\,s^{-1}})\,t]$. The diffusion radii for representative particle energies are listed in Tab.~\ref{diff table}.

Given the characteristic molecular-cloud diameter $2R \sim 10\,\mathrm{pc}$ (see Sec.~\ref{density}), it is plausible that these high-energy particles diffuse through the cloud and generate the observed extended $\gamma$-ray emission, 4FGL~J2220.8+6319e.

\begin{table}
\centering
\caption{The ranges of diffusion radius (in units of pc) for particles with different energies, for diffusion times ranging from $10^{4}$ to $10^{5}\ \mathrm{yr}$.}
\label{diff table}
\begin{tabular}{lccc}
\hline
Energy & 1GeV & 10GeV & 85.76GeV \\
$R_{\mathrm{diff,p}}$(pc) & 2.1 - 7.7 & 3.7 - 13.8 & 6.3 - 23.6 \\
\hline
Energy & 1GeV & 10GeV & 127.73GeV \\
$R_{\mathrm{diff,e}}$(pc) & 2.0 - 6.5 & 3.6 - 11.5 & 6.9 - 21.9 \\
\hline
\end{tabular}
\end{table}

\subsection{Comparison with known (candidate) protostellar jet--$\gamma$-ray systems}
\begin{figure*}  
    \centering
    \includegraphics[width=\textwidth]{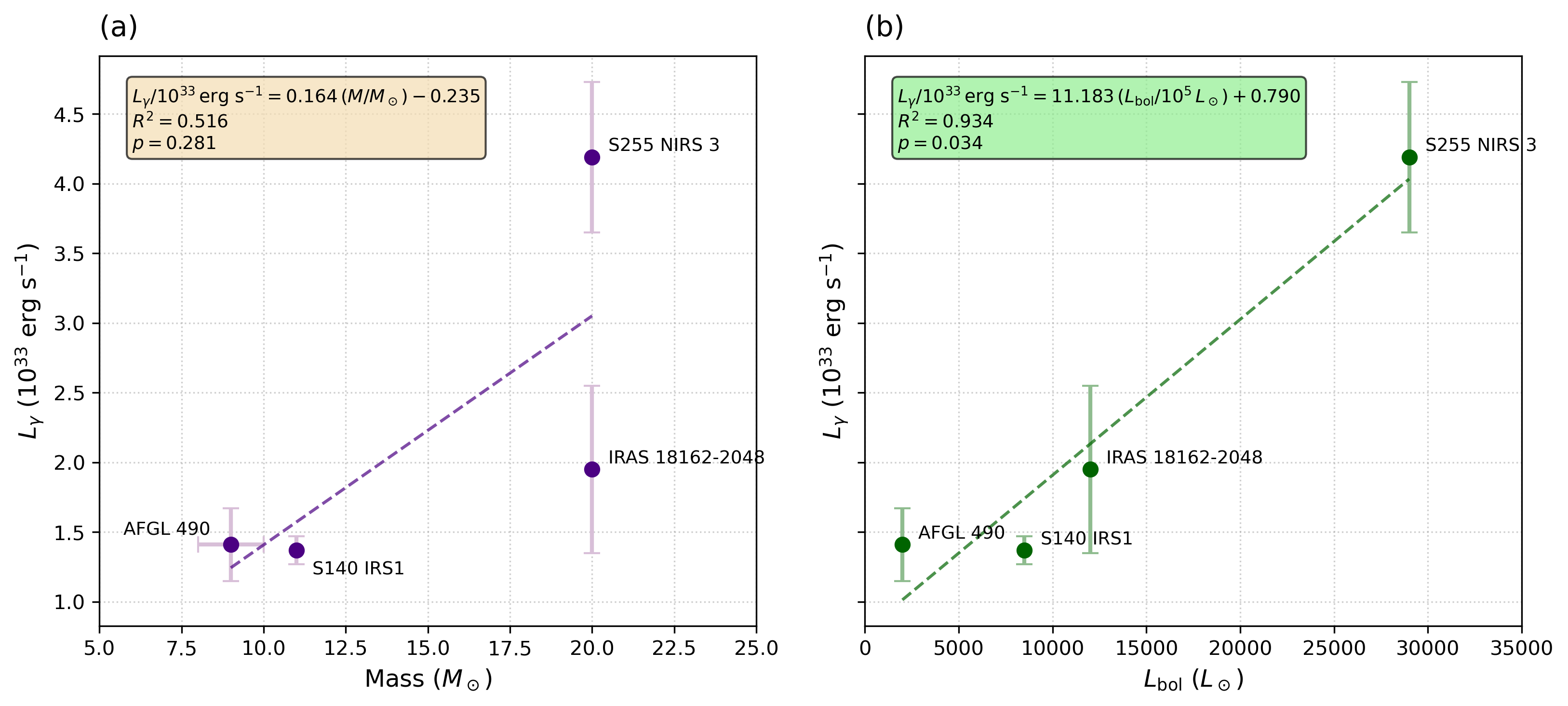}
    \caption{Physical properties of four protostellar jet systems (possibly) associated with $\gamma$-ray emission. The corresponding parameters are given in Tab.~\ref{tab:gamma_bol_ratio}.}
    \label{protostar compare}
\end{figure*}

Finally, we compile the known (or candidate) cases of $\gamma$-ray emission from protostellar jets reported so far (see Sec.~\ref{intro}) and compare them with S140 IRS1; the results are summarised in Tab.~\ref{tab:gamma_bol_ratio} and Fig.~\ref{protostar compare}. A notable quantity is $L_\gamma/L_{\rm bol}$ ($\times 10^{5}$): HH~80-81, S255~NIRS3, and S140~IRS1 all exhibit values around 4, whereas AFGL~490 shows a significantly higher value of $\sim$18. 

A linear regression analysis reveals that \(L_\gamma\) exhibits no significant correlation with the source mass (\(R^2 = 0.516\), \(p = 0.281\)). In contrast, a tight positive correlation is found between \(L_\gamma\) and \(L_{\mathrm{bol}}\) (\(R^2 = 0.934\), \(p = 0.034\)), suggesting that the \(\gamma\)-ray luminosity primarily scales with the overall energy output rather than the final stellar mass. However, given the extremely limited sample size of only four sources, this preliminary correlation is statistically fragile and cannot be generalized as a universal physical law at present. Dedicated observations targeting protostellar objects across a wider range of bolometric luminosities are necessary to robustly verify this physical association. 

To gain further physical insight into the underlying mechanisms, we can decompose this ratio using the following expression:
\begin{equation}
    \frac{L_{\gamma}}{L_{\mathrm{bol}}} = \left( \frac{L_{\mathrm{acc}}}{L_{\mathrm{bol}}} \right) \times \left( \frac{L_{\mathrm{jet}}}{L_{\mathrm{acc}}} \right) \times \left( \frac{L_{\gamma}}{L_{\mathrm{jet}}} \right)
\end{equation}
where the term \(L_{\mathrm{acc}}/L_{\mathrm{bol}}\) characterizes the accretion intensity of massive protostars, representing the proportion of bolometric luminosity driven by ongoing accretion. The main difficulty in estimating this ratio stems from the considerable uncertainty in determining \(L_{\mathrm{acc}}\). Among the three known \(\gamma\)-ray emitting protostellar systems mentioned above, this ratio can only be evaluated for AFGL 490 (\(\sim 42.5\%\))~\citep{2026MNRAS.549ag896Y}, while it reaches \(\sim 62.4\%\) for S140 IRS1. For IRAS 18162--2048 (HH 80-81), no estimate of \(L_{\mathrm{acc}}\) is available, and for S255~NIRS~3, only the \(L_{\mathrm{acc}}\) during the outburst phase is known, rather than in quiescence. Next, the term \(L_{\mathrm{jet}}/L_{\mathrm{acc}}\) describes the fraction of accretion energy converted into the kinetic power of the jet, potentially associated with certain magnetohydrodynamic (MHD) processes. For example, \citet{2007LNP...723...21C}\ reported \(L_{\mathrm{jet}}/L_{\mathrm{acc}} \sim 7.5\%\), which is the value adopted in this work. Finally, the term \(L_{\gamma}/L_{\mathrm{jet}}\) represents the efficiency with which the jet's kinetic energy is converted into \(\gamma\)-ray luminosity, which currently stands as the parameter about which we have the most limited understanding.

\renewcommand{\arraystretch}{1.2}
\begin{table*}
    \centering
    \caption{Physical properties of four protostellar jet systems (possibly) associated with $\gamma$-ray emission. Here, $L_\gamma$ denotes the total $\gamma$-ray luminosity in the $0.1$--$500$~GeV band, computed from the spectral parameters and considering only their uncertainties (distance uncertainties are not included). The parameter references for the individual sources are: HH~80-81 \citep{2025AA...695A..11M}, S255~NIRS~3 \citep{2023MNRAS.523..105D}, and AFGL~490 \citep{2026MNRAS.549ag896Y}.}
    \label{tab:gamma_bol_ratio}
    \begin{tabular}{lcccc}
        \hline
        Protostellar System & $L_\gamma$ ($10^{33}$\,erg\,s$^{-1}$) & Mass ($M_\odot$) & $L_{\rm bol}$ ($L_\odot$) & $L_\gamma/L_{\rm bol}$ ($\times 10^{5}$) \\
        \hline
        IRAS 18162$-$2048 (HH 80$-$81) & $1.95 \pm 0.60$ & $20$ & $12000$ & $4.25 \pm 1.31$ \\
        S255 NIRS 3 & $4.19 \pm 0.54$ & $20$ & $29000$ (quiescent) & $3.77 \pm 0.49$ \\
        AFGL 490 & $1.41 \pm 0.26$ & $8$--$10$ & $2000$ & $18.42 \pm 3.40$ \\
        S140 IRS1 & $1.37 \pm 0.10$ & $11$ & $8500$ & $4.21 \pm 0.31$ \\
        \hline
    \end{tabular}
\end{table*}

\section{CONCLUSION}
\label{conclusion}
In this paper, we investigate the GeV $\gamma$-ray emission from the S140 region. We confirm that the region is best modelled as an extended source, 4FGL~J2220.8+6319e, with a TS value of 1029.9 ($\sim 32.1\sigma$). Its spectrum follows a log-parabola shape, described by the parameters $\alpha = 2.44 \pm 0.05$, $\beta = 0.29 \pm 0.03$, and $E_b = 1200$~MeV.

Given the absence of any other $\gamma$-ray counterpart within the region, a plausible scenario is that the observed emission arises from particles accelerated by the jet of the massive young stellar object S140~IRS1, producing $\gamma$-rays via hadronic and/or leptonic processes. Meanwhile, a comparison across available protostellar jet systems reveals a linear tendency between $L_\gamma$ and $L_{\rm bol}$ ($R^2 = 0.934$, $p = 0.034$). Nonetheless, due to the extreme sparsity of the current sample (four sources in total), this correlation should be interpreted with caution as a preliminary conjecture rather than a definitive conclusion.

\begin{acknowledgments}
We acknowledge the generous support of the Fermi Large Area Telescope Collaboration and the Fermi Science Support Center (FSSC) for developing and providing the publicly available data, software, and analysis tools. We are also grateful to the anonymous referee for the constructive comments that have significantly improved the quality of this manuscript.
\end{acknowledgments} 


\bibliographystyle{aasjournalv7}
\bibliography{example}   

\end{document}